%% file: Main.tex
\documentclass[twocolumn]{aastex631}
\usepackage{booktabs}
\usepackage{amsmath}
\usepackage{graphicx}
\usepackage{subfigure}
\usepackage{longtable}
\usepackage{xspace}

\shorttitle{}
\shortauthors{Taamoli et al.}
\graphicspath{{./}{figures/}}

\newcommand{\photz}{photo-\textit{z}\xspace}

\begin{document}
\title{The Role of the Cosmic Web in Galaxy Evolution Across $0.4<z<4$}

\input{authors}

\begin{abstract}
We investigate the dependence of stellar mass and star formation activity on the topology of hosting large-scale structures using mass-complete galaxy samples in four redshift bins spanning from $z = 0.4$ to $z = 4$ using the COSMOS2020 catalog. Cosmic web components (clusters, filaments, and field) are identified by applying the Multiscale Morphology Filter to galaxy density maps, and group memberships (central, satellite, isolated) are assigned to each galaxy by the Friends of Friends (FoF) algorithm. Out to $z < 1$, we find that stellar mass increases toward denser structures, driven primarily by central and satellite galaxies, with differences up to $\sim0.35$ dex between field and clusters. At these redshifts, star formation rate (SFR) is suppressed by $\sim 0.1$ dex in clusters compared to field galaxies, particularly for centrals, while the SFR of star-forming galaxies shows a much weaker dependence on the type of host structure, indicating that the quenched population primarily drives the observed environmental dependence. Specific SFR (sSFR) shows the strongest decline for centrals in clusters at $z < 1$, consistent with enhanced quenching in dense environments. At $1 < z < 2$, star formation activity becomes largely independent of environment, suggesting a transitional regime in the role of large-scale structure. At $z > 2$, we observe a mild increase in SFR ($\sim+0.1$ dex) and sSFR ($\sim+0.13$ dex) from field to cluster galaxies, pointing to a possible reversal of environmental trends during early cosmic epochs. These results highlight the evolving influence of environment on galaxy evolution and set the stage for deeper investigations with upcoming wide-field surveys.
\end{abstract}


\section{Introduction}\label{sec:introduction}
% Opening: Cosmic Web and Galaxy Environments
The Cosmic Web, made up of voids, sheets, filaments, and clusters, forms through the gravitational collapse of the initial density perturbations \citep{bond1996} and plays a significant role in the evolution of galaxies. Matter flows from underdense regions into dense nodes along filaments, creating a network that leads to galaxy and halo formation. This intricate network of structures, predicted by simulations and recently confirmed by observations, regulates the process of galaxy formation and evolution \citep{Springel2005,Scoville2007,cautun2014,ansari2019}.

% Galaxy Environments
Galaxy properties show clear trends with environment, particularly at lower redshifts ($z<1$). Galaxies in clusters tend to be massive, quiescent, and composed mostly of older stellar populations, while those in the field are often less massive, star-forming, and disk-like. \citep{dressler1980, balogh2004,kauffmann2004,baldry2006,Bamford2009,peng2010,woo2013}. Environmental effects have shown their signature in star formation \citep{Scoville2013,darvish2016,chartab2020,taamoli2024}, metal enrichment and gas content \citep{sattari2021,chartab2021B}, and morphological properties \citep{Mandelbaum2006,capak2007,Bamford2009,Laigle2025}. The observed trends are stronger at lower redshifts ($z<1$), where galaxies have had more time to evolve within their host structures \citep{peng2010,Poggianti2017,chartab2020,taamoli2024,zheng2024}.

% Galaxy Groups
As intermediate-scale structures bridging galaxies and large-scale structures (hereafter \emph{LSS}), galaxy groups form the fundamental building blocks of the cosmic web \citep{Eke2004,Toni2025}. Since most galaxies reside in groups, they represent the typical dense environments for understanding quenching mechanisms \citep{McGee2009,Yang2007}. The internal structure of groups, especially the distinction between central and satellite galaxies, plays a critical role in shaping galaxy properties \citep{vandenBosch2008,Knobel2015}. Centrals typically reside at the bottom of the potential well and may retain access to cold gas, although this depends on halo mass and redshift, as virial shocks and heating can suppress accretion in massive halos at low redshift. Satellites are more susceptible to environmental processes such as ram-pressure stripping and starvation that suppress star formation \citep{Wetzel2013,Peng2012}. Groups can merge into clusters over time, making them crucial in understanding the hierarchical buildup of dense environments.

% Cosmic Web
On scales beyond groups and clusters, galaxies are part of the cosmic web, a network of large structures that influences their evolution. Simulations predict that galaxy properties vary across the cosmic web, highlighting the importance of the topology of these structures beyond local density alone \citep{Springel2005}. Efforts to detect and characterize the cosmic web in both simulations and observations have made progress using techniques such as analysis of density field curvature (Hessian matrix) \citep{Romero2009, cautun2014,libeskind2018,Malavasi2020}. However, identifying the cosmic web in observational data remains challenging due to projection effects, survey limitations, and the complex, scale-dependent nature of the cosmic web geometry \citep{Kraljic2018}.

% Cosmic Web quenching
The different components of the cosmic web are expected to host distinct physical processes that influence galaxy evolution. Clusters are dominated by high-density, virialized halos, where mechanisms such as ram-pressure stripping and galaxy harassment act to suppress star formation \citep{gunn_infall_1972,moore_galaxy_1996}. Filaments are less dense, elongated structures that can also drive galaxy transformations through environmental pre-processing, regulation of gas accretion, and mild interactions \citep{Bahe2013,Alpaslan2016}. Observational studies have found enhanced quiescent fractions in filaments compared to the field, indicating that environmental quenching can begin well before galaxies enter cluster cores \citep{darvish2016,Malavasi2017}. Recent Euclid observations further show that galaxy morphology and orientation vary with position relative to cosmic filaments \citep{Laigle2025}. Understanding these differences is essential for linking galaxy properties to their large-scale environments. In particular, filament intersections and the connectivity of dense nodes are important for studying the formation and growth of clusters and proto-clusters \citep{Matsuda2005,Koyama2013,Daikuhara2025,Gouin2025}.

% Cosmic web in observations
Recent advances in deep, wide-area photometric surveys now allow us to reconstruct the density field across redshift and identify different components of the cosmic web, despite the presence of projection systematics \citep{Laigle2016,darvish2017Cosmicweb,Malavasi2025Euclid}.
% and compare them with predictions from theoretical models and simulations \citep{Kraljic2018}
The COSMOS field, in particular, provides an ideal dataset for cosmic web studies, due to its deep, high-quality photometric coverage over a wide contiguous area \citep{Scoville2007,Laigle2016,weaver2022}. This enables studies on how galaxy properties vary across the cosmic web and over cosmic time. Despite these advances, identifying cosmic web components observationally remains non-trivial, due to persistent systematic uncertainties and survey limitations \citep{Kraljic2018}. Additionally, variations in method-such as differences in density estimators or structure-finding algorithms-can lead to inconsistent classifications \citep{libeskind2018}.

% This work
In this work, we investigate how star formation activity and quenching relate to large-scale structures. Using 135 redshift slices from $0.4<z<4$, we build on previously constructed density fields \citep{taamoli2024} to identify clusters and filaments through the analysis of the density field Hessian matrix. Galaxies are classified as field, filament, or cluster systems based on the topology of their surrounding environment. Because many quenching processes act at the halo scale and can dilute trends with large-scale environment, we also analyze galaxy groups to separate halo-driven effects from the cosmic web context. We further assign galaxies to groups using a friends-of-friends algorithm and identify centrals, satellites, and isolated systems based on stellar mass ranking within groups. This study offers new observational insight into how the cosmic web shapes galaxy evolution across cosmic time.

% Outline
The paper is organized as follows. In Section \ref{sec:data}, we briefly review the reconstruction of the large-scale structure maps at different redshifts in the COSMOS field and our selection criteria for galaxy samples. In Section \ref{sec:methods}, we describe the methods used to identify cosmic web components and galaxy groups. Section \ref{sec:results} presents our results on the evolution of galaxies within different large-scale environments. Finally, in Section \ref{sec:summary}, we summarize the main findings of this work.

Throughout this work, we assume a flat $\Lambda\text{CDM}$ cosmology with $H_{0} = 70 \, \text{km} \, s^{-1} \text{Mpc}^{-1}$, $\Omega_{m0} = 0.3$, and $\Omega_{\Lambda0}= 0.7 $. All magnitudes are expressed in the AB system \citep{Oke1983}, and the physical parameters are measured assuming a Chabrier initial mass function \citep{Chabrier2003IMF}.

\section{Data} \label{sec:data}
We use the COSMOS2020 photometric \citep{weaver2022} catalogue based on \texttt{The Farmer} photometry \citep{weaver_farmer_2023} with the physical parameters derived from the \texttt{LePHARE}\footnote{\href{https://www.cfht.hawaii.edu/~arnouts/LEPHARE/lephare.html}{https://www.cfht.hawaii.edu/\(\sim\)arnouts/LEPHARE/lephare.html}} SED fitting code \citep{ilbert2006, Arnouts2002lephare}. For environmental densities, we rely on the catalog from \citet{taamoli2024}, which uses photometric redshifts (hereafter \photz) and their probability distributions (zPDF) from the \texttt{Farmer+Lephare} combination to produce a density catalog for 210,621 galaxies at \(0.4 < z < 5\) using adaptive Kernel Density Estimation (wKDE) method. Following the procedure in \citet{ilbert2013}, galaxies are classified as star-forming or quiescent using the rest-frame \texttt{NUV-r} versus \texttt{r-J} color-color diagram, which helps distinguish quiescent galaxies from dust-reddened star-forming systems \citep{Williams2009,Patel2012}. The selection criteria for this sample and density field construction process are explained in detail in \citet{taamoli2024}.

For stellar masses and sSFRs used in the following sections, we use median values of likelihood functions for these parameters reported in COSMOS2020 (``\texttt{lp\_mass\_med}'', ``\texttt{lp\_SFR\_med}'', and ``\texttt{lp\_sSFR\_med}'').

\section{Methods} \label{sec:methods}
In this section, we briefly describe the methods used to identify Cosmic Web structures and galaxy groups. The full methodology is outlined in detail in \citealt{Aragon2007} and \citealt{darvish2017Cosmicweb}. For consistency and completeness, we provide a brief overview and highlight the key steps, including the parameters adopted and modifications we applied in our method to extract cosmic web components.

\subsection{Cosmic Web Extraction} \label{sec:cosmicweb}
To extract \emph{LSS} components of various sizes and shapes from the 2D density fields, we apply the Multi-scale Morphology Filter (MMF) algorithm \citep{Aragon2007}. This method relies on analyzing the Hessian matrix of the density field:

\begin{equation}
    H(\Vec{r}) = 
\begin{bmatrix}
\nabla^2_{xx} \sigma (\Vec{r}) & \nabla^2_{xy} \sigma (\Vec{r}) \\
\nabla^2_{yx} \sigma (\Vec{r}) & \nabla^2_{yy} \sigma (\Vec{r})
\end{bmatrix}
\end{equation}

where $\nabla_{ij}^{2}\sigma(\Vec{r})$ denotes the second-order directional derivative of the density field. Given the multi-scale nature of the problem, we compute the Hessian matrix at different smoothing scales, calculate the cluster and filament signals for each galaxy at each scale, and assign to each galaxy the maximum signal across all scales. For a physical smoothing scale $L$, the elements of the Hessian matrix are computed as:

\begin{equation}
H_{L}(\vec{r}) = \int 
\frac{(x_i - x_i')(x_j - x_j') - \delta_{ij} L^2}{L^4} 
\, G_{L}(\vec{r}', \vec{r}) \sigma(\vec{r}') \, d\vec{r}'  
\end{equation}

where $G_{L}(\vec{r}', \vec{r})$ is a 2D Gaussian smoothing kernel. Physically, $H_L(\vec{r})$ measures the local curvature of the density field after smoothing on scale $L$. The convolution with the second derivatives of the Gaussian filter quantifies how rapidly the density changes along different directions around each position. In this work, we use the physical scales $L = 0.5, 0.7, 1, 1.5, 2, 3, 4$ Mpc (see \cite{Aragon2007} for the rationale behind this choice). If a region in the density field has a cluster-like topology, both eigenvalues of the Hessian matrix will be negative. In the case of filamentary structures, the eigenvalue corresponding to the eigenvector perpendicular to the filament axis is expected to be negative and have a larger absolute value than the one along the filament. Assuming $\lambda_{1}$ and $\lambda_{2}$ are the eigenvalues with $|\lambda_{1}| \geq |\lambda_{2}|$, a morphology mask $\varepsilon$, is defined as:

\begin{equation}
\begin{aligned}
Clusters: \varepsilon_{\text{cl}}(\vec{r}) &= 
\begin{cases}
0 & \text{if } \lambda_{1} > 0 \ \text{or} \ \lambda_{2} > 0, \\
1 & \text{otherwise},
\end{cases} \\[1ex]
Filaments: \varepsilon_{\text{fi}}(\vec{r}) &= 
\begin{cases}
0 & \text{if } \lambda_{2} > 0, \\
1 & \text{otherwise}.
\end{cases}
\end{aligned}
\end{equation}

The degree to which the density field resembles a cluster or filament is quantified using a resemblance function $D$:

\begin{equation}
D_{\text{cl}}(\vec{r}) = \frac{|\lambda_{2}|}{|\lambda_{1}|}, 
\qquad 
D_{\text{fi}}(\vec{r}) = 1 - \frac{|\lambda_{2}|}{|\lambda_{1}|}
\end{equation}
Both resemblance functions range from 0 to 1. For clusters, $D_{\mathrm{cl}}\simeq1$ indicates comparable curvature along both principal directions, as expected for a compact, approximately isotropic structure, while $D_{\mathrm{cl}}\simeq0$ indicates strongly unequal curvatures. Conversely, $D_{\mathrm{fi}}\simeq1$ corresponds to strong curvature in one direction and weak curvature in the other, as expected for a filament, whereas $D_{\mathrm{fi}}\simeq0$ indicates similar curvature along both directions.
Following \citet{Frangi1998}, we define an $M$ function that acts as a morphological filter, enhancing features consistent with cluster or filament shapes while suppressing noise. This function provides a measure of how likely a region is to be classified as a cluster or filament:

\begin{equation}
M_{\text{(cl/fi)}}(\vec{r}) = \exp\!\left(-\frac{D_{\text{(fi/cl)}}(\vec{r})}{2\beta^{2}}\right)
\end{equation}

Where $\beta$ regulates the aggressiveness of this function. We adopt $\beta = 0.5$, following \citep{Frangi1998,Aragon2007,darvish2017Cosmicweb}. To reduce the effect of background noise, we include a contrast function based on the norm of the Hessian eigenvalues, which reflects the overall strength of curvature in the field:

\begin{equation}
I(\vec{r}) = 1 - \exp\!\left(-\frac{\text{Norm}^{2}}{2c^{2}}\right), 
\quad 
\text{Norm} = \sqrt{\lambda_{1}^{2} + \lambda_{2}^{2}}
\end{equation}

Here, $c$ sets the sensitivity of the response. Following \citet{Frangi1998}, we set $c = 0.5 \times \max(\text{Norm})$ in each redshift slice. The final signal at smoothing scale $L$, $S_L$, is computed as the convolution of the three parameters:

\begin{equation}
S_{L}(\vec{r}) = \varepsilon(\vec{r}) \otimes M(\vec{r}) \otimes I(\vec{r})
\end{equation}

and the final signal map, at each redshift slice, is constructed by taking the maximum signal across all scales at each point:
\begin{equation}\label{eq:signal}
S(\vec{r}) = \max(S_{L}(\vec{r}))
\end{equation}

The final step is to calculate the cluster or filament signal for each galaxy by summing the signal values at the galaxy’s position across all redshift slices using a weighted approach:

\begin{equation}
S_{\text{cl/fi}}(G) = \sum_{i} w^g_{i} S^i_{\text{cl/fi}}(\vec{r})
\end{equation}

Here, $S_{\text{cl/fi}}(G)$ is the cluster or filament signal assigned to galaxy $G$, $w_i^g$ is the weight of galaxy $g$ in the $i$th redshift slice derived from its photometric redshift PDF, and $S^i_{\text{cl/fi}}(\vec{r})$ is the cluster or filament signal at the galaxy’s position in the $i$th slice, as defined in Equation~\ref{eq:signal}.

\begin{figure}
    \centering
    \includegraphics[width = 1\linewidth]{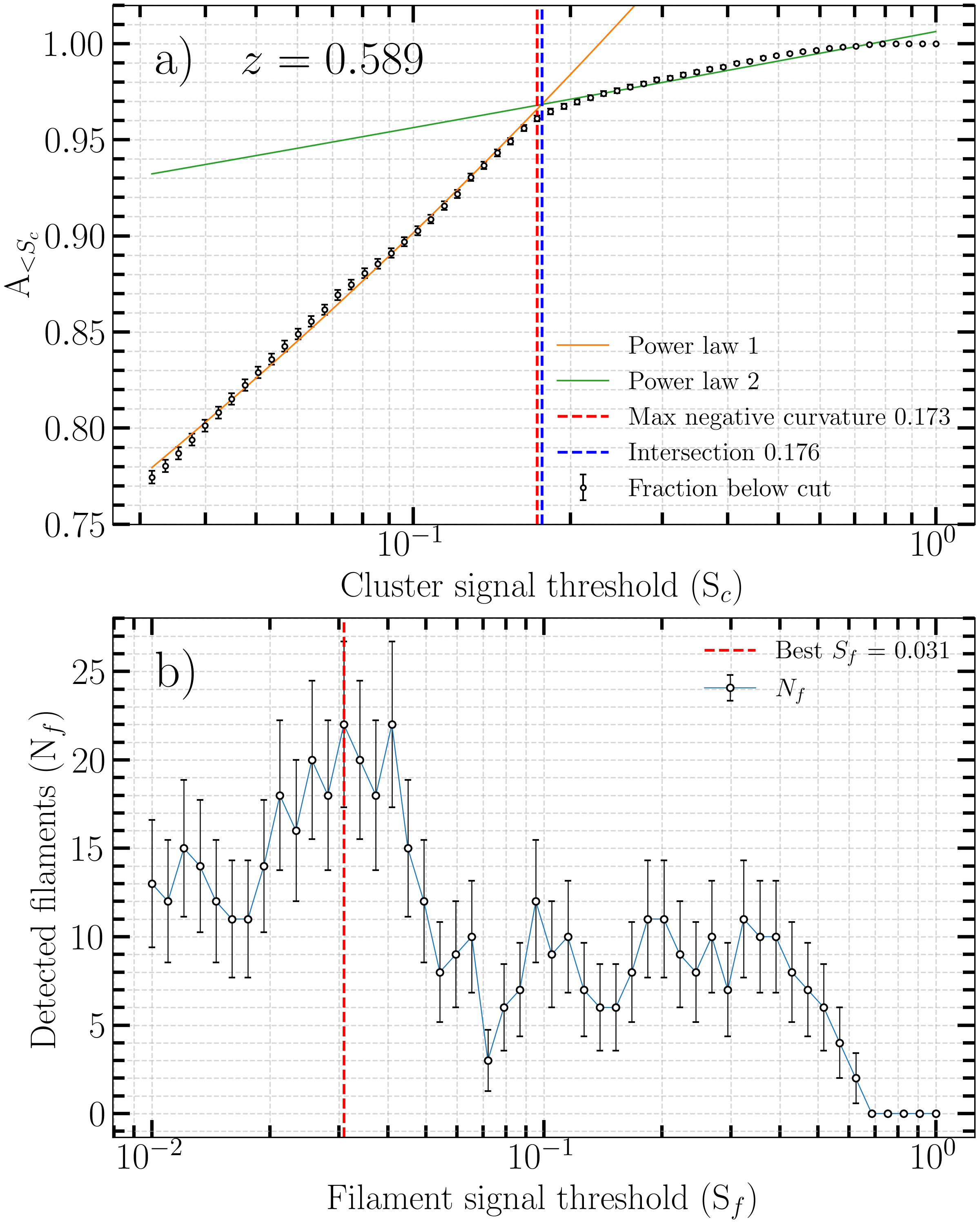}
    \caption{Determination of the optimal cluster and filament signal thresholds for a redshift slice at $z = 0.589$. \textbf{(a)} The cumulative area fraction below a given cluster signal threshold, $A_{<S_c}$, is shown as a function of $S_c$ (black circles). The curve is fit with two power-law functions (orange and green lines), and their intersection is shown as a blue dashed line ($S_c = 0.176$). In this work, the location of maximum negative curvature (red dashed line) is adopted as the criterion for cluster detection. \textbf{(b)} The number of detected filaments, $N_f$, is plotted as a function of the filament signal threshold, $S_f$. The best filament threshold is selected as the value that maximizes $N_f$, shown by the red dashed line at $S_f = 0.031$. Error bars represent binomial/Poisson uncertainties in panel a/b. These thresholds are used to classify galaxies into cluster, filament, and field environments in this redshift slice.}
    \label{fig:signalscut}
\end{figure}

\begin{figure*}
    \centering
    \includegraphics[width = 1\textwidth]{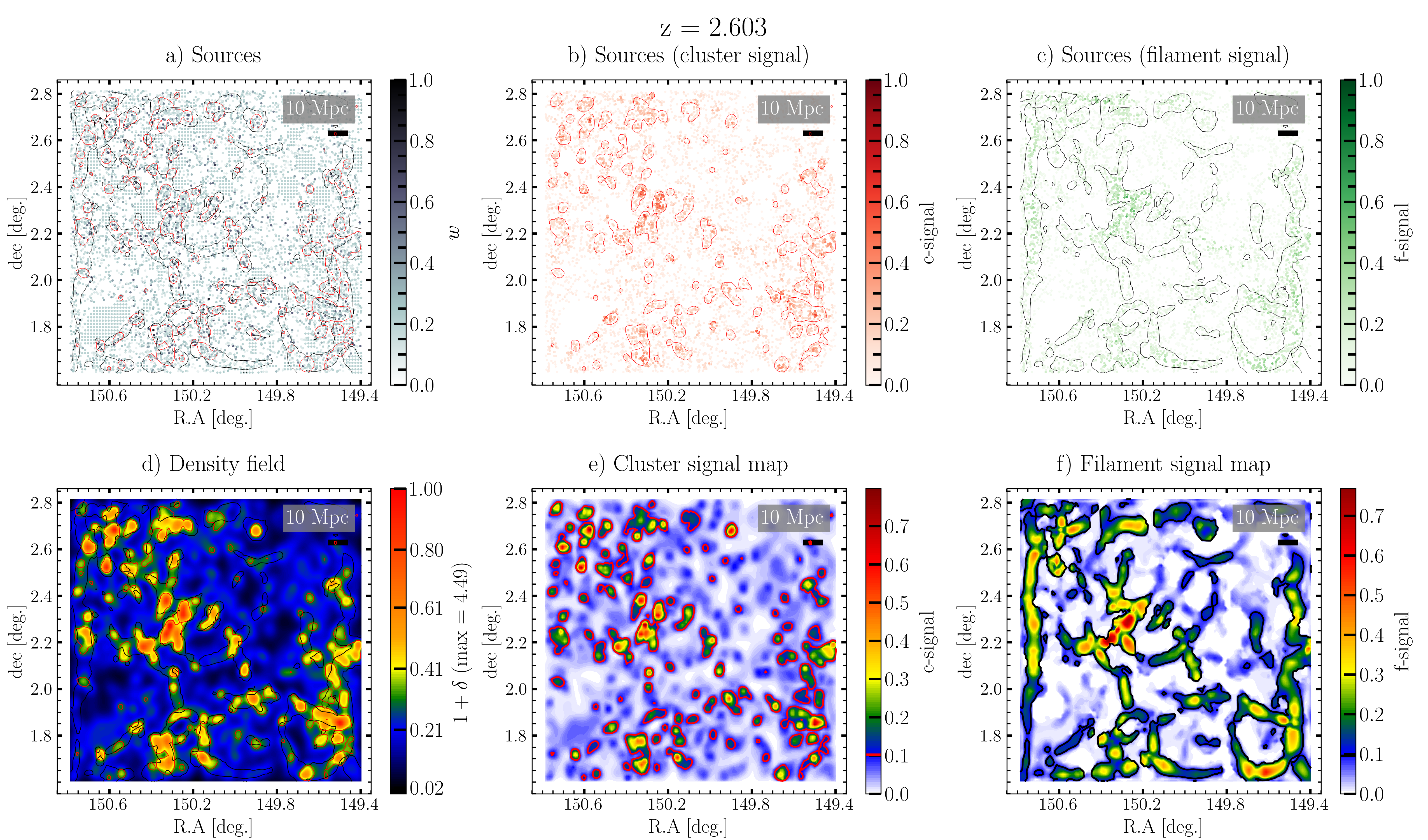}
    \caption{This figure illustrates the process of extracting cosmic web components. The top row shows the spatial distribution of galaxies in the redshift slice at $z = 2.603$, color-coded by galaxies' weight (integral of zPDF within the redshift slice) (a), cluster signal (b), and filament signal (c). The bottom row displays the corresponding continuous density field (d), cluster signal map (e), and filament signal map (f). Contours in panels (b), (c), (e), and (f) indicate the cluster and filament signal thresholds described in Section \ref{sec:cosmicweb}. For details on panels (a) and (d), see \citep{taamoli2024}.}
    \label{fig:signals}
\end{figure*}

Having final cluster/filament signals for all galaxies, the next step is to decide whether a galaxy should be classified as a cluster, filament, or field galaxy. For each signal, we define a threshold based on visual inspection of the signal maps and the redshift evolution of the signal metrics. We selected thresholds that preserved coherent, extended structures while avoiding the fragmentation of filaments and cluster regions at high thresholds and the excessive merging of structures with the surrounding field at low thresholds. For clusters, we track the area fraction below a given cluster signal and identify the point where the curvature of this function becomes most negative, indicating that the area fraction becomes less sensitive to increasing signal thresholds. This point is determined for each redshift slice and fit with a linear relation to define the redshift-dependent cluster threshold: $S_c(z) = -0.013z + 0.14$. An alternative approach adopted by \citealt{darvish2017Cosmicweb} defines the threshold as the intersection of two power-law fits to this function. Figure~\ref{fig:signalscut} illustrates the process for a representative redshift slice at $z = 0.589$. Panel (a) shows the cumulative area fraction $A_{<S_c}$ as a function of the cluster signal threshold $S_c$ (black circles), along with power-law fits (orange and green lines) and their intersection line (blue dashed line), and the location of the maximum negative curvature (red dashed line). While the two methods produce similar results in most redshift slices, in cases where they diverge, the curvature-based threshold adopted here more closely matches the structures seen in the signal maps.

For filaments, we compute the number of detected filamentary structures $N_f$ as a function of filament signal threshold $S_f$. The number of filaments increases with $S_f$ up to a certain value, after which it declines due to fragmentation or merging of structures. We adopt the peak of this distribution as the best filament threshold and fit it across slices with a linear relation: $S_f (z) = -0.02z + 0.15$. Panel (b) in Figure \ref{fig:signalscut} shows the number of detected filaments $N_f$ versus $S_f$, with error bars representing the Poisson uncertainties. The red dashed line marks the filament threshold corresponding to the maximum number of distinct filamentary components detected.

We classify galaxies as cluster, filament, or field galaxies based on how their cluster and filament signals compare to these thresholds: a galaxy is classified as a cluster galaxy if its cluster signal is greater than or equal to the cluster threshold at that redshift, and the cluster signal is also greater than or equal to its filament signal. Among the remaining galaxies, those with filament signal greater than or equal to the filament threshold and also greater than or equal to their cluster signal are classified as filament galaxies. The rest are labeled as field galaxies \citep{darvish2017Cosmicweb}:
\begin{equation}
% \text{Galaxy type} =
\begin{cases}
\text{Cluster} & \text{if } S_c \geq \max(S_f,\, S_c^{\text{th}}) \\
\text{Filament} & \text{if } S_f \geq \max(S_c,\, S_f^{\text{th}}) \\
\text{Field} & \text{otherwise}
\end{cases}
\end{equation}

Here, $S_c$ and $S_f$ are the cluster and filament signal values assigned to each galaxy, while $S_c^{\text{th}}$ and $S_f^{\text{th}}$ are the corresponding signal thresholds determined for each redshift slice.

Figure~\ref{fig:signals} presents the process and results of the cosmic web extraction pipeline for a redshift slice at $z = 2.603$. The top panels show the spatial distribution of galaxies, color-coded by their weight (panel a), cluster signal (panel b), and filament signal (panel c). These signals reflect the degree of morphological resemblance to clusters or filaments at the position of each galaxy within this redshift slice. In the bottom row, panel (d) shows the smoothed 2D density field, while panels (e) and (f) present the resulting cluster and filament signal maps, respectively. The contours overlaid on panels (b), (c), (e), and (f) mark the significance thresholds used to define cluster-like and filament-like regions. Further details about the construction of the galaxy  (panel a) and the smoothed density maps (panel d) can be found in \citep{taamoli2024}.

\subsection{Identification of Galaxy Groups}\label{sec:groups}
To identify galaxy groups within our sample, we use the Friends-of-Friends (FoF) algorithm \citep{Eke2004FoF,darvish2017Cosmicweb}, which links pairs of galaxies based on their projected and line-of-sight separations. Two galaxies $i$ and $j$ with redshifts $z_i$ and $z_j$, and angular separation $\Delta \theta_{ij}$, are considered linked if they satisfy the following conditions:
\begin{equation}
D_c(z) \Delta \theta_{ij} \leq b_\perp  n(z)^{-1/2}, \quad |z_i - z_j| \leq \alpha \sigma_{\Delta z/(1+z)}
\end{equation}
Here, $D_c(z)$ is the comoving distance at the mean redshift of the pair, $n(z)$ is the median galaxy number density at redshift $z$, $b_\perp$ is the transverse linking length in units of the mean projected separation, and $\alpha$ sets the linking scale along the line of sight based on the photometric redshift uncertainty $\sigma_{\Delta z/(1+z)}$. We adopt $b_\perp = 0.4$ and $\alpha = 1.5$, which provide a reasonable balance between breaking up groups at small scales and over-percolating structures at large scales. After group identification, the most massive galaxy in each group is designated as the ``central'', while the remaining members are classified as ``satellites''. Galaxies not assigned to any group are labeled as ``isolated''. This classification allows us to examine galaxy properties in relation to both group membership and hosting \emph{LSS} environment.

\subsection{Mass complete subsamples}\label{sec:CWEBmasscompsubsamps}
In each redshift bin \([z_{\min},\,z_{\max})\) we define a mass–complete subsample. For every galaxy, we estimate the stellar mass it would have at the survey \(K_s\) flux limit. For each redshift bin, we then set the mass-completeness threshold as the 90th percentile of this limiting mass, computed separately for each galaxy subsample we compare (the overall and star-forming populations), and retain only galaxies with \(M_\star\) above the corresponding threshold. Because this threshold does not depend on cosmic-web environment, within each subsample the same cut is applied uniformly to its field, filament, and cluster galaxies.

\begin{figure}
    \centering
    \includegraphics[width = 1\linewidth]{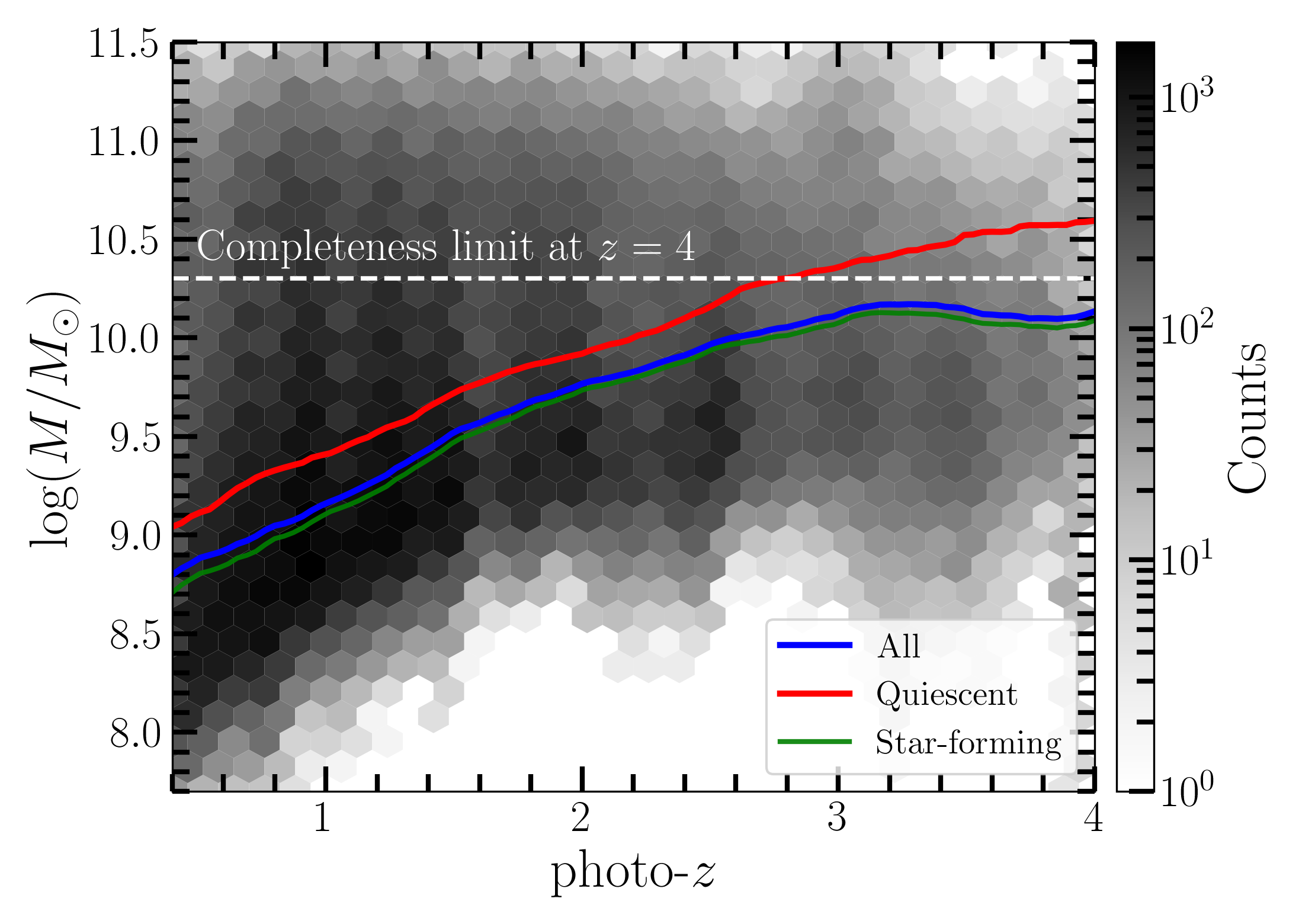}
    \caption{Gray map shows galaxy number density in the $\big(z_{\mathrm{phot}}, \log(M/M_\odot)\big)$ plane. Solid curves give the stellar mass completeness derived in sliding windows of redshift using the 90th percentile of the per galaxy limiting mass. Completeness limits are shown for All galaxies (blue), Quiescent (red), and Star forming (green). The horizontal dashed line marks the global quiescent completeness floor set by the highest redshift bin near $z=4$. The colorbar indicates counts on a logarithmic scale. The quiescent curve lies above the others because quiescent systems have higher $M_\star/L$ in $K_s$.}
    \label{fig:CWCompleteness}
\end{figure}

This per-bin procedure reduces SED–dependent selection effects and enables fair comparisons across environments within each redshift bin \citep{pozeti2010,ilbert2013}. Figure~\ref{fig:CWCompleteness} shows the galaxy distribution in the \((z_{\mathrm{phot}},\,\log(M/M_\odot))\) plane. Solid curves trace the redshift–dependent stellar–mass completeness: blue for the full sample, red for quiescent galaxies, and green for star–forming galaxies. The horizontal dashed line marks the global quiescent completeness floor ($\log(M/M_\odot)\sim 10.3$), set by the highest redshift bin \(2\le z<4\); which is a bin-averaged threshold.

\begin{table*}[t]
\centering
\small
\caption{Cosmic Web Environments and Group Membership properties for a sub-sample of galaxies.}
\vspace{-0.1in}
\begin{tabular}{llllllcllcccc}
% \hline\hline
\noalign{\hrule height 0.5pt}
\noalign{\vskip 0.7pt}
\noalign{\hrule height 0.5pt}
ID & Photo-$z$ & $\alpha_{2000}$ & $\delta_{2000}$ & density & Physical & SF/Q\textsuperscript{a} & Cluster & Filament & Cosmic Web & Group& Group\\
\noalign{\vskip -0.8ex}
& & (deg) & (deg) & contrast & density && signal & signal & flag\textsuperscript{b} & flag\textsuperscript{c}& ID\\
\hline
54  & 0.5910 & 150.722986 & 2.543470 &  0.5938 & 0.0969 & 1 & 0.0192 & 0.1056 & 0 & 1 & 3236 \\
57  & 0.9672 & 150.718184 & 1.792733 &  0.0924 & 0.0987 & 0 & 0.0418 & 0.2422 & 1 & 0 & 0 \\
73  & 0.4290 & 149.468025 & 1.625609 &  0.7569 & 0.1164 & 1 & 0.0402 & 0.1800 & 1 & 1 & 603 \\
74  & 1.5087 & 150.488035 & 1.875422 &  0.1076 & 0.1101 & 1 & 0.0003 & 0.0685 & 0 & 1 & 23941 \\
85  & 1.0800 & 150.476436 & 2.334178 &  0.2600 & 0.0904 & 1 & 0.0137 & 0.0980 & 0 & 0 & 0 \\
91  & 3.1054 & 150.476213 & 2.334883 & -0.0736 & 0.0895 & 0 & 0.0716 & 0.0604 & 0 & 2 & 35853 \\
97  & 1.9183 & 150.479121 & 1.875552 &  0.0101 & 0.0768 & 1 & 0.0069 & 0.0293 & 0 & 2 & 28356\\
\hline
\end{tabular}
\label{table:cwebproperties}
\vspace{-0.13in}
\parbox{0.95\linewidth}{\flushleft \textbf{Note}\\
\textsuperscript{a} 0: quiescent, 1: star-forming\\
\textsuperscript{b} 0: field, 1: filament, 2: cluster\\
\textsuperscript{c} 0: isolated, 1: satellite, 2: central\\
This table is available in its entirety in machine-readable form.}
\end{table*}

\section{Results and Discussion} \label{sec:results}
Table~\ref{table:cwebproperties} presents COSMOS IDs, photometric redshifts, and coordinates from the COSMOS2020 catalog, along with environmental density values from \citet{taamoli2024}, cosmic web classification-including cluster and filament signals- and an environment flag indicating whether a galaxy resides in the field (0), filament (1), or cluster (2), and group membership, labeled as isolated (0), satellite (1), or central (2) for a subsample of galaxies. The remainder of this section analyzes galaxy properties as a function of their cosmic web environment and group membership. The full electronic catalog for 210621 galaxies in our sample is published in its entirety.

\subsection{Overall Statistics}\label{sec:overalStats}

\begin{figure}
    \centering
    \includegraphics[width = 1\linewidth]{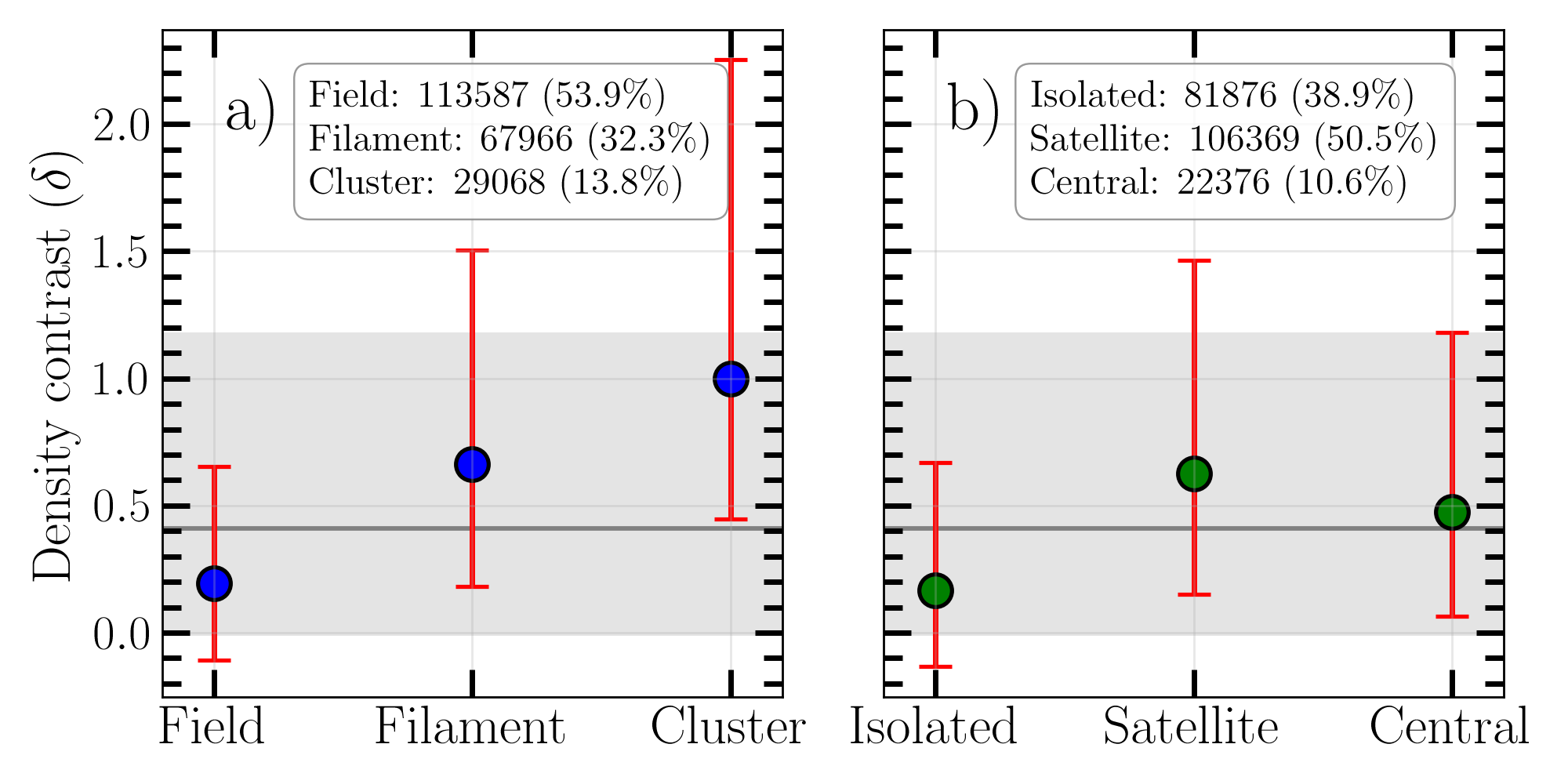}
    \caption{Median galaxy density contrast ($\delta$) in different environments. (a) Galaxies classified according to their large-scale structure membership: field, filament, and cluster. (b) Galaxies classified according to group membership: isolated, satellites, and centrals. Symbols indicate the median density contrast for each class, with error bars showing the 16th–84th percentile range of the distributions. The horizontal line and shaded band represent the median density contrast of the full sample ($\delta \approx 0.41$) and its 16th–84th percentile range. Annotation boxes report the number and fraction of galaxies in each category.}
    \label{fig:overalStats}
\end{figure}

Figure \ref{fig:overalStats} summarizes the median of environmental density contrast for our galaxy sample across both \emph{LSS} environments and group membership. In panel (a), cluster galaxies have the highest environmental density with a median density contrast of $\delta = 1.00^{+1.26}_{-0.55}$, followed by filament galaxies with $\delta = 0.66^{+0.84}_{-0.48}$, while field galaxies reside in the lowest-density environments with $\delta = 0.19^{+0.46}_{-0.30}$. Panel (b) presents the same statistics for group membership: satellites are found in the densest regions with $\delta = 0.63^{+0.84}_{-0.47}$, centrals occupy somewhat lower densities with $\delta = 0.48^{+0.70}_{-0.41}$, and isolated galaxies lie in the least dense environments with $\delta = 0.17^{+0.50}_{-0.30}$. The large error bars, however, should be considered when interpreting these central values. For comparison, the median of the full sample is $\delta = 0.41^{+0.77}_{-0.42}$, shown by the horizontal line and shaded region in both panels. The error bars and shaded bands correspond to the 16th–84th percentile range of the distributions, highlighting the broad spread of environments spanned by each class. The population fractions reported in the annotation boxes emphasize that the majority of galaxies in our sample reside in lower-density environments, such as filaments and fields, rather than clusters.

We examined the distribution of stellar mass across different \emph{LSS} environments for a sample of galaxies above the universal completeness limit ($\log(M/M_\odot)\sim 10.3$), set for the highest redshift bin ($2\le z<4$) of this work (horizontal dashed line in Figure~\ref{fig:CWCompleteness}). We find that field galaxies, while residing in the lowest-density regions, contain the largest share of the total stellar mass in our sample, contributing $\sim$51.1\%. Filaments account for $\sim$32.1\% of the stellar mass, while cluster galaxies host only $\sim$16.8\% of the total stellar mass. This trend reflects the fact that clusters represent a relatively minor share of the overall stellar mass budget. In contrast, the large number of galaxies in the field results in a dominant contribution to the total stellar mass, even though these galaxies typically reside in underdense regions.

\begin{figure*}
    \centering
    \includegraphics[width = 0.8\textwidth]{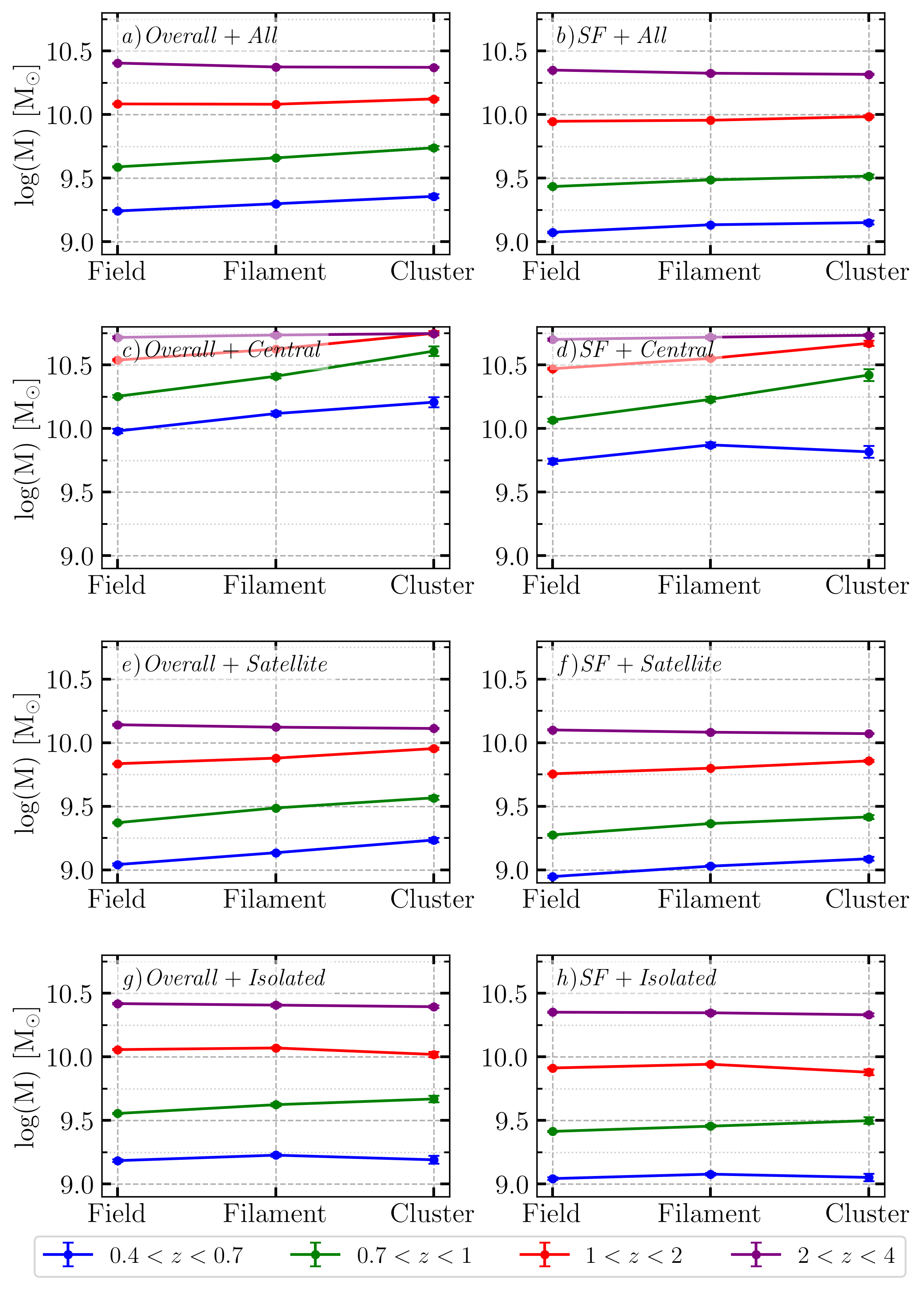}
    \caption{Median stellar mass ($\log(M_\ast/\mathrm{M}_\odot)$) as a function of cosmic web environment (field, filament, cluster), shown across four redshift bins: $0.4 < z < 0.7$ (blue), $0.7 < z < 1$ (green), $1 < z < 2$ (red), and $2 < z < 4$ (purple). Error bars represent the standard error on the median. Left and right columns correspond to the full galaxy population (star-forming + quiescent) and star-forming galaxies only, respectively. The first row includes all galaxies regardless of group membership, while the second, third, and fourth rows show central, satellite, and isolated galaxies, respectively.}
    \label{fig:CWEB-SM}
\end{figure*}

Our stellar mass fractions can be directly compared with those reported by \citet{darvish2017Cosmicweb} in the COSMOS field. They found that galaxies with $\log(M/M_\odot)\geq9.6$, and within the range of $0.1<z<1.2$, contribute 14.3\%, 41.1\%, and 44.6\% of the stellar mass in clusters, filaments, and the field, respectively. While our cluster fraction of 16.8\% is consistent with their reported value, we find a larger fraction in the field (51.1\% versus 44.6\%) and a correspondingly smaller share in filaments (32.1\% versus 41.1\%). The main trade-off arises between filament and field galaxies, although we should note that the redshift cuts differ between these two works. Sensitivity tests show that adjusting the pipeline parameters primarily shifts galaxies between these two classes, while the identification of clusters remains comparatively stable. This behavior reflects a broader challenge: filaments, due to their lower densities and complex morphologies, are intrinsically more difficult to identify and are therefore more susceptible to being classified as field regions \citep{cautun2014,libeskind2018}. These differences arise from the selected parameters and method variations, including differences in sample selection (e.g., mass completeness limit), redshift slicing, and cosmic web identification, but the overall picture remains consistent: clusters, though occupying the smallest volume, host a non-negligible stellar mass fraction, while the bulk of the stellar mass resides in filaments and the field. Another source of discrepancy can be attributed to the different redshift ranges probed. Our comparison suggests that at higher redshifts ($z\gtrsim1$), a larger fraction of stellar mass resides in the field relative to filaments, consistent with the expected evolution of galaxy clustering over cosmic time.

\subsection{Physical Properties Across Different Environments}\label{sec:PhysicalProperties}
In this section, we analyze the physical properties of galaxies in mass-complete samples across four redshift bins spanning $0.4<z<4$, examining their dependence on \emph{LSS} environment and group membership.

\subsubsection{Stellar Mass}
Figure~\ref{fig:CWEB-SM} shows the median stellar mass ($\log(M_\ast/\mathrm{M}_\odot)$) of galaxies as a function of cosmic web environment (field, filament, cluster), separated into four redshift bins: $0.4 < z < 0.7$ (blue), $0.7 < z < 1$ (green), $1 < z < 2$ (red), and $2 < z < 4$ (purple). Error bars represent the standard error of the median value. The left column includes the full galaxy population (star-forming and quiescent), while the right column is restricted to star-forming galaxies. Rows correspond to different galaxy types: the first row presents the whole sample regardless of group membership, followed by centrals, satellites, and isolated galaxies in the second, third, and fourth rows, respectively.

Across the full sample (panel a), median stellar mass increases modestly from field to cluster environments, with the strongest increase observed at $z<1$, particularly at $0.7 < z < 1$ (green), rising from $\log(M_\ast/\mathrm{M}_\odot) \sim 9.59$ in the field to $\log(M_\ast/\mathrm{M}_\odot) \sim 9.74$ in clusters (a $\sim0.15$ dex increase). Centrals (panel c) drive this trend: at $0.7 < z < 1$, their median mass increases from $\log(M_\ast/\mathrm{M}_\odot) \sim 10.25$ in the field to $\log(M_\ast/\mathrm{M}_\odot) \sim 10.61$ in clusters ($\sim 0.36$ dex), the strongest environmental dependence observed in the figure, and at $0.4 < z < 0.7$ from $\log(M_\ast/\mathrm{M}_\odot) \sim 10.00$ to $\log(M_\ast/\mathrm{M}_\odot) \sim 10.23$ ($\sim 0.23$ dex). Star-forming centrals (panel d) also show a strong environmental trend, particularly at $0.7 < z < 1$, where the median rises from the field to clusters (by $\sim 0.35$ dex). Satellites (panels e–f) exhibit slightly weaker environmental dependence with a $ \sim 0.2$ dex increase between field and cluster environments at $0.7 < z < 1$. Isolated galaxies (panels g–h) show essentially flat trends, with stellar masses remaining nearly constant across environments at all redshifts.

The right column demonstrates that when only star-forming galaxies are considered, environmental differences become weaker overall. For instance, in the star-forming sample (panel b), stellar masses vary by less than 0.1 dex across environments at all redshifts. The suppression of environmental trends among star-forming galaxies highlights that the most significant mass segregation occurs within the quiescent population, which has completed the mass assembly process. At higher redshifts ($z > 2$), each subsample shows nearly constant median stellar mass across environments (negligible environmental dependence), with absolute levels of $\log(M_\ast/\mathrm{M}_\odot) \sim 10.4$ for the full population, $\sim 10.7$ for centrals, and $\sim 10.1$ for satellites.

In summary, the most significant environmental dependence of stellar mass is observed among central and satellite galaxies at $z < 1$, where median masses increase by up to $\sim0.35$ dex from field to cluster environments. Isolated galaxies show much weaker or negligible variations, and the trend weakens generally when restricting the analysis to star-forming galaxies. This indicates that the observed stellar mass–environment relation is primarily driven by quiescent central and satellite galaxies in dense regions at intermediate redshifts.

\begin{figure*}
    \centering
    \includegraphics[width = 0.8\textwidth]{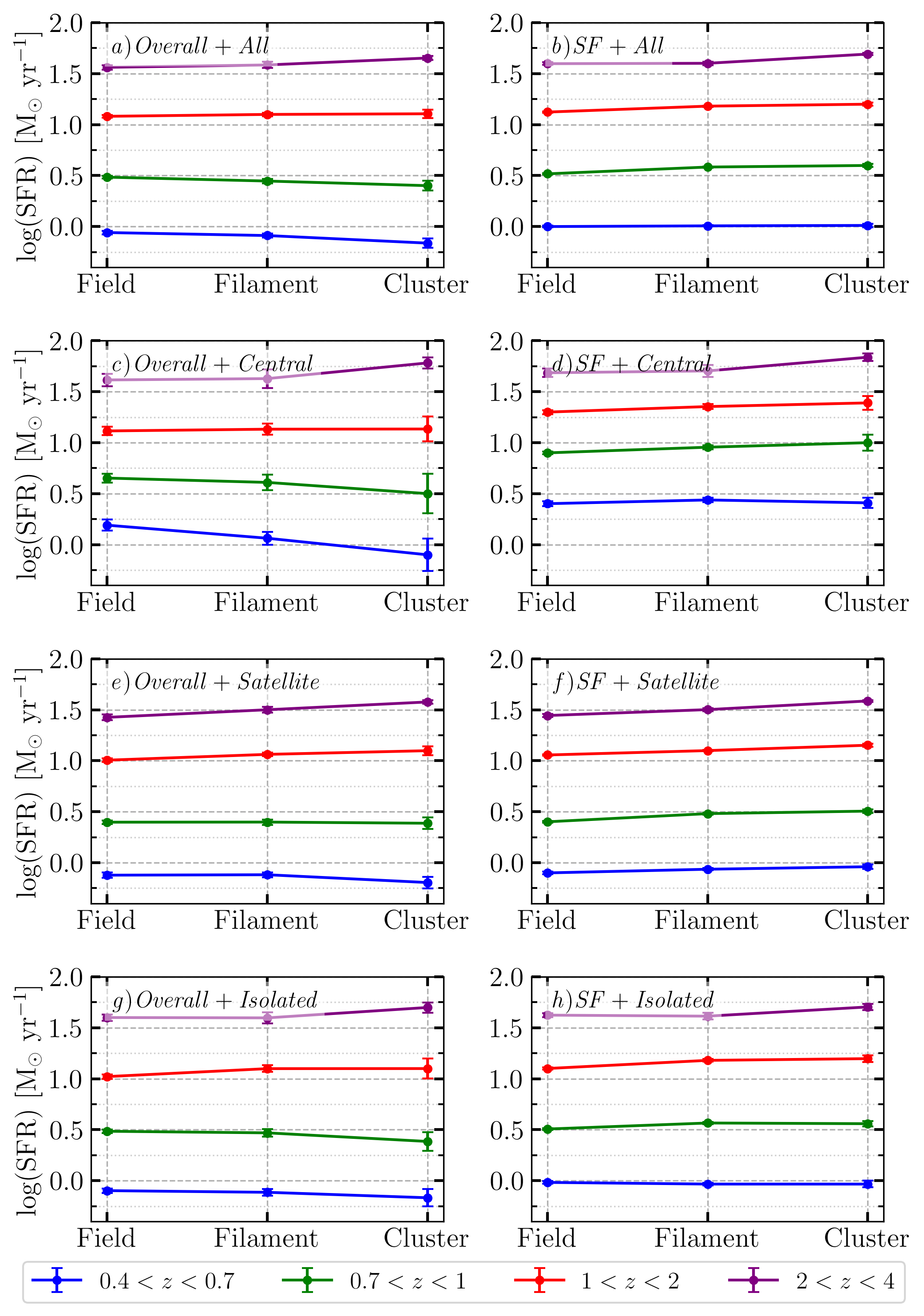}
    \caption{Median star formation rate ($\log(\mathrm{SFR}/\mathrm{M}_\odot,\mathrm{yr}^{-1})$) as a function of cosmic web environment (field, filament, cluster), shown across four redshift bins: $0.4 < z < 0.7$ (blue), $0.7 < z < 1$ (green), $1 < z < 2$ (red), and $2 < z < 4$ (purple). Error bars represent the standard error of the median. Left and right columns correspond to the full galaxy population (star-forming + quiescent) and star-forming galaxies only, respectively. The first row includes all galaxies regardless of group membership, while the second, third, and fourth rows show central, satellite, and isolated galaxies, respectively.}
    \label{fig:CWEB-SFR}
\end{figure*}

\subsubsection{Star Formation Rate (SFR)}
Figure \ref{fig:CWEB-SFR} presents the median star formation rate ($\log \mathrm{SFR}[\mathrm{M}_\odot \mathrm{yr}^{-1}]$) of galaxies as a function of cosmic web environment (field, filament, cluster) across four redshift bins: $0.4 < z < 0.7$ (blue), $0.7 < z < 1$ (green), $1 < z < 2$ (red), and $2 < z < 4$ (purple). Error bars denote the standard error on the median. The left column shows results for the full galaxy population (star-forming + quiescent), while the right column is restricted to star-forming galaxies. Rows correspond to all galaxies (top), centrals (second), satellites (third), and isolated galaxies (bottom).

At high redshift ($2 < z < 4$), galaxies display the highest SFRs, with median values around $1.6$–$1.65$ in the \textit{Overall+All} sample, dropping to $\sim 1.4-1.6$ for satellites. There is a mild but increasing environmental variation from field to cluster ($ \sim 0.1$ dex), for the overall and star-forming samples. An almost flat behavior is seen for the star-forming population at $1 < z < 2$, where SFR remains nearly constant across field, filament, and cluster environments ($\sim1.1$–$1.2$). This shows that at this cosmic epoch, star formation activity is not primarily governed by the \emph{LSS} environment, suggesting that this redshift range can be considered a transitional epoch between $z<1$ and $z>2$ regimes.

Clearer environmental differences emerge at lower redshift ($0.4<z<1$). In the overall galaxy sample at $0.4 < z < 1$ (blue and green), median SFR declines by $\sim -0.1$ dex from field to clusters. This suppression is more pronounced for centrals and satellites and at the lowest redshifts (blue): centrals fall by $\sim -0.3$  dex, and satellites decrease by $\sim 0.1$ dex from field to cluster (panels c \& e). This decrease can be seen for isolated galaxies, although shallower than for centrals and satellites. In contrast, the corresponding star-forming subsamples show nearly flat trends in general, although the increase from field to cluster (by $\sim 0.1$ dex) persists at $z>2$ for all group categories.

Overall, the most significant environmental dependence of SFR is seen at $z < 1$, where galaxies in clusters exhibit reduced star formation compared to those in filaments and the field. This suppression is most pronounced in the overall population and particularly for centrals, while satellites and isolated galaxies remain less affected. The weakening of the trends in the star-forming-only panels, particularly at $z<2$, highlights that the observed environmental differences are primarily due to the rising quenched fraction in clusters at late times.

\subsubsection{Specific Star Formation Rate (sSFR)}
\begin{figure*}
    \centering
    \includegraphics[width = 0.8\textwidth]{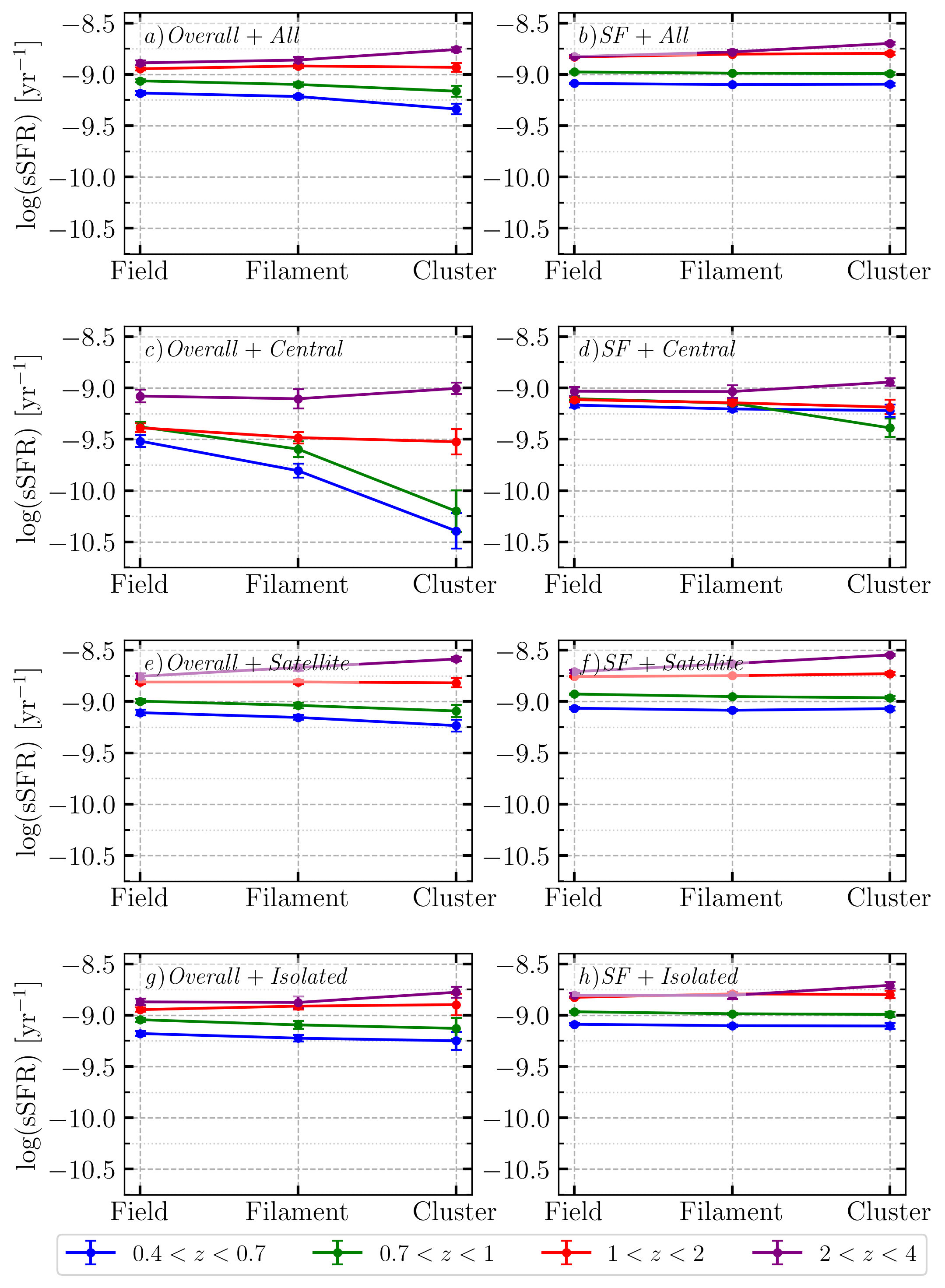}
    \caption{Median specific star formation rate ($\log(\mathrm{sSFR}/\mathrm{yr}^{-1})$) as a function of cosmic web environment (field, filament, cluster), shown across four redshift bins: $0.4 < z < 0.7$ (blue), $0.7 < z < 1$ (green), $1 < z < 2$ (red), and $2 < z < 4$ (purple). Error bars represent the standard error on the median. Left and right columns correspond to the full galaxy population (star-forming + quiescent) and star-forming galaxies only, respectively. The first row includes all galaxies regardless of group membership, while the second, third, and fourth rows show central, satellite, and isolated galaxies, respectively.}
    \label{fig:CWEB-sSFR}
\end{figure*}

Figure~\ref{fig:CWEB-sSFR} shows the median specific star formation rate as a function of \emph{LSS} environment type, across redshift and group membership. At $1<z<2$, sSFR values remain nearly flat in the \textit{Overall+All} sample (panel a), with small environmental dependence. Differences become more evident at lower redshift ($0.4<z<1$), particularly for central galaxies (panels a and c), where sSFR decreases toward clusters, indicating stronger quenching in dense environments. Satellites and isolated galaxies show weaker variations with \emph{LSS} type, and when restricting to the star-forming population, the environmental trends largely disappear. This again confirms that the observed decline in sSFR at low redshift is mainly driven by the higher quenched fraction of centrals in clusters.

\section{Summary} \label{sec:summary}

We investigated the dependence of stellar mass, star formation rate (SFR), and specific SFR (sSFR) on cosmic web environment using mass-complete subsamples of galaxies in four redshift bins spanning $0.4 < z < 4$, drawn from the COSMOS2020 catalog. Building on the density measurements presented in \cite{taamoli2024}, we derived \emph{LSS} characteristics, such as cluster and filament signals, and assigned group memberships (central, satellite, isolated) to galaxies in our sample. This framework enabled us to trace the evolution of galaxy physical properties across cosmic time and in diverse environments.

The main results can be summarized as follows:

\begin{enumerate}
    \item \textbf{Overall statistics:} Cluster and central/satellite galaxies occupy the densest regions, while field/filament and isolated galaxies reside in the most diffuse environments. Despite their low densities, field galaxies contribute the largest share of the total stellar mass ($\sim$51.1\%), followed by filaments ($\sim$32.1\%) and clusters ($\sim$16.8\%).
    
    \item \textbf{Stellar mass–environment:} Median stellar mass increases from field to cluster at $z < 1$, driven mainly by central galaxies, which show up to $\sim0.35$ dex higher masses in clusters compared to the field. Satellites show weaker trends ($\sim0.2$ dex), while isolated galaxies exhibit a nearly flat relation with \emph{LSS} type. This trend is stronger in the overall sample consisting of quiescent and star-forming galaxies and weaker in the star-forming population, particularly for isolated galaxies.

    \item \textbf{SFR–environment:} At $z < 1$, SFR is suppressed in clusters relative to the field, with declines of up to $\sim0.3$ dex for centrals, while satellites and isolated galaxies show much weaker suppression ($\lesssim 0.1$ dex), driven by a higher quenched fraction. At $1 < z < 2$, SFR shows shallower environmental dependence, marking a transitional regime. At $z > 2$, SFRs are highest overall, with a mild increase ($\sim0.1$ dex) from field to cluster environments, even among star-forming galaxies.

    \item \textbf{sSFR–environment:} At $z < 1$, sSFR declines toward cluster environments, most strongly for central galaxies, indicating enhanced quenching for central galaxies in dense regions. Satellites and isolated galaxies show weaker trends, and environmental differences largely vanish among star-forming galaxies.

    \item \textbf{Role of quiescent population:} The environmental dependence of stellar mass, SFR, and sSFR is primarily driven by the quiescent population. When restricting to star-forming galaxies, trends in stellar mass and SFR largely disappear, indicating that the observed environmental effects arise from an increased quenched fraction in dense environments rather than shifts in the star-forming main sequence.

    \item \textbf{Evolution of central, satellite, isolated galaxies:} Central and satellite galaxies exhibit the strongest environmental trends in stellar mass and SFR at $z < 1$, reflecting accelerated environmental mechanisms for these galaxies. In contrast, isolated galaxies remain largely unaffected by environment across all redshifts, with nearly flat trends in both mass and star formation. At higher redshifts ($z > 2$), differences between group populations diminish, suggesting a convergence in galaxy properties across different environments during earlier cosmic epochs.

\end{enumerate}

Our findings suggest that the connection between galaxy properties and the cosmic web is primarily driven by the buildup of quiescent galaxies in dense environments at late times. While we observe a redshift evolution in these trends, with a possible transitional phase at $1 < z < 2$ and indications of trend reversal at higher redshifts ($z>2$), the results at early epochs remain limited by observational constraints, including the quality of photometric data, uncertainties in SED-derived physical properties, and the depth of current surveys \citep{chartab2025latis}. With the advent of current and upcoming deep and wide-area surveys such as \textit{Euclid}, the Vera C. Rubin Observatory, and the Nancy Grace Roman Space Telescope, it will become possible to extend this analysis to larger volumes, higher redshifts, and a broader range of environments. Future work can refine this picture by improving photometric redshift accuracy, incorporating additional galaxy properties such as morphology and AGN activity, and linking observations with hydrodynamical simulations to better constrain the environmental drivers of galaxy evolution. These efforts will enable a more complete understanding of how galaxies co-evolve with the cosmic web across cosmic time.

\section{Acknowledgments}
% We are grateful to the anonymous referee for their helpful comments that greatly improved the quality of this work. 

% Some of the data used in this study were obtained at the W.M. Keck Observatory, which operates as a scientific partnership among the California Institute of Technology, the University of California, and the National Aeronautics and Space Administration. The Observatory was made possible by the generous financial support of the W.M. Keck Foundation. The authors wish to acknowledge the profound cultural role of Maunakea's summit within the indigenous Hawaiian community. 
This work is based on data products from observations made with ESO Telescopes at the La Silla Paranal Observatory under ESO program ID 179.A-2005 and on data products produced by CALET and the Cambridge Astronomy Survey Unit on behalf of the UltraVISTA consortium. This work is based in part on observations made with the NASA/ESA Hubble Space Telescope, obtained from the Data Archive at the Space Telescope Science Institute, which is operated by the Association of Universities for Research in Astronomy, Inc., under NASA contract NAS 5-26555. B.M. and S.T. acknowledge support from the National Science Foundation under Grant No. 2206813. J.R.W. acknowledges that support for this work was provided by The Brinson Foundation through a Brinson Prize Fellowship grant.
\bibliography{Ref}{}
\bibliographystyle{aasjournal}
\end{document}

%% file: authors.tex
\author[0000-0003-0749-4667]{Sina Taamoli}
\affiliation{Department of Physics and Astronomy, University of California, Riverside, 900 University Ave, Riverside, CA 92521, USA}
\email{taamolisina@gmail.com}

\author[0000-0001-5846-4404]{Bahram Mobasher}
\affiliation{Department of Physics and Astronomy, University of California, Riverside, 900 University Ave, Riverside, CA 92521, USA}

\author[0009-0007-3673-4523]{Hossein Hatamnia}
\affiliation{Department of Physics and Astronomy, University of California, Riverside, 900 University Avenue, Riverside, CA 92521, USA}

\author[0000-0003-3691-937X]{Nima Chartab}
\affiliation{Infrared Processing and Analysis Center, California Institute of Technology, Pasadena, CA 91125, USA}

\author[0000-0003-1614-196X]{John R. Weaver}\thanks{Brinson Prize Fellow}
\affiliation{MIT Kavli Institute for Astrophysics and Space Research, 70 Vassar Street, Cambridge, MA 02139, USA}
% \email{john.weaver.astro@gmail.com}

\author[0000-0002-0236-919X]{Ghassem Gozaliasl}
\affiliation{Department of Computer Science, Aalto University, P.O. Box 15400, FI-00076 Espoo, Finland}
\affiliation{Department of Physics, University of Helsinki, P.O. Box 64, FI-00014 Helsinki, Finland}

\author[0000-0002-6610-2048]{Anton M. Koekemoer}
\affiliation{Space Telescope Science Institute, 3700 San Martin Drive,
Baltimore, MD 21218, USA}

\author[0000-0002-0930-6466]{Caitlin M. Casey}
\affiliation{Department of Physics, University of California, Santa Barbara, Santa Barbara, CA 93106, USA}
\affiliation{Cosmic Dawn Center (DAWN), Denmark}

\author[0000-0001-9187-3605]{Jeyhan S. Kartaltepe}
\affiliation{Laboratory for Multiwavelength Astrophysics, School of Physics and Astronomy, Rochester Institute of Technology, 84 Lomb Memorial Drive, Rochester, NY
14623, USA}

\author[0009-0006-3071-7143]{Aryana Haghjoo}
\affiliation{Department of Physics and Astronomy, University of California, Riverside, 900 University Avenue, Riverside, CA 92521, USA}

% \author[0000-0003-3578-6843]{Peter L. Capak}
% \affiliation{Infrared Processing and Analysis Center, California Institute of Technology, Pasadena, CA 91125, USA}

% \author[0009-0003-3097-6733]{Ali Hadi}
% \email{ahadi005@ucr.edu}
% \affiliation{Department of Physics and Astronomy, University of California, Riverside, 900 University Avenue, Riverside, CA 92521, USA}

\author[0000-0002-7303-4397]{Olivier Ilbert}
\affiliation{Aix Marseille Univ, CNRS, LAM, Laboratoire d’Astrophysique de Marseille, Marseille, France}

\author[0000-0002-9489-7765]{Henry J. McCracken}
\affiliation{Institut d’Astrophysique de Paris, UMR 7095, CNRS, and Sorbonne Université, 98 bis boulevard Arago, F-75014 Paris, France}

\author[0000-0002-1233-9998]{David B. Sanders}
\affiliation{Institute for Astronomy, University of Hawaii, 2680 Woodlawn Drive, Honolulu, HI 96822, USA}

\author[0000-0002-0438-3323]{Nicholas Scoville}
\affiliation{California Institute of Technology, 1200 E. California Boulevard, Pasadena, CA 91125, USA}

\author[0000-0003-3631-7176]{Sune Toft}
\affiliation{Cosmic Dawn Center (DAWN), Denmark}
\affiliation{Niels Bohr Institute, University of Copenhagen, Jagtvej 128, DK-2200 Copenhagen, Denmark}